
\magnification=\magstephalf
\hoffset=0.0 true cm
\voffset=1.0 true cm
\vsize=24.0 true cm
\hsize=17.0 true cm
\baselineskip=12pt
\parskip=5pt
\parindent=22pt
\raggedbottom

\font\medrm=cmr10 scaled \magstep1
\font\bigbf=cmb10 scaled \magstep2

\def\pp{\noindent\parshape 2 0.0 truecm 17.0 truecm 0.5 truecm 16.5 truecm}

\def\etal{{\frenchspacing\it et al.} }
\def\lsim{\hbox{ \rlap{\raise 0.425ex\hbox{$<$}}\lower 0.65ex\hbox{$\sim$} }}
\def\gsim{\hbox{ \rlap{\raise 0.425ex\hbox{$>$}}\lower 0.65ex\hbox{$\sim$} }}


\parskip=0pt
\noindent To appear in: {\sl Applications of Digital Image Processing XX}

\noindent ed. A. Tescher, {\it Proc. S.P.I.E.} {\bf 3164}, in press (1997)
\parskip=5pt

\vskip 2.0 truecm

\centerline{\bigbf Data-Mining a Large Digital Sky Survey:}
\medskip

\centerline{\bigbf From the Challenges to the Scientific Results}
\bigskip

\centerline{\medrm S.G.~Djorgovski$^a$, R.R.~de Carvalho$^{a,c}$,
S.C.~Odewahn$^a$, R.R.~Gal$^a$, }
\medskip
\centerline{\medrm J.~Roden$^b$, P.~Stolorz$^b$, and A.~Gray$^b$}
\medskip
\medskip

\centerline{\medrm $^a$ Palomar Observatory, Caltech, Pasadena, CA 91125, USA}
\medskip

\centerline{\medrm $^b$ Jet Propulsion Laboratory, Pasadena, CA 91109, USA}
\medskip

\centerline{\medrm $^c$ Observatorio Nacional, CNPq, Rio de Janeiro, Brasil}
\medskip

\bigskip
\bigskip
\centerline{\bf Abstract}
\medskip

The analysis and an efficient scientific exploration of the Digital Palomar
Observatory Sky Survey (DPOSS) represents a major technical challenge.  The
input data set consists of 3 Terabytes of pixel information, and contains
a few billion sources.  We describe some of the specific scientific problems
posed by the data, including searches for distant quasars and clusters of
galaxies, and the data-mining techniques we are exploring in 
addressing them.  Machine-assisted discovery methods may become essential for
the analysis of such multi-Terabyte data sets.  New and future approaches
involve unsupervised classification and clustering analysis in the Giga-object
data space, including various Bayesian techniques.  In addition to the searches
for known types of objects in this data base, these techniques may also offer 
the possibility of discovering previously unknown, rare types of astronomical
objects. 

\medskip
\centerline{{\sl Keywords:}~~ 
data mining, sky surveys, clustering analysis, unsupervised classification }
\medskip
\medskip

\bigskip
\centerline{\bf 1.~ Introduction: The Challenge of Terabyte Data Sets}
\medskip

The problem of the data glut has arrived, in virtually every field of science
and technology.  However, raw data, no matter how expensively obtained, are of
limited utility without the effective ability to process them quickly and
thoroughly, and to refine the essence of scientific knowledge from them.  Are 
we ready to exploit what the new era of nearly unlimited scientific information
has to offer? 

The motivation and goals behind our work are to confront the problem of
extracting interesting scientific results from vast amounts of digital
data in an efficient, yet statistically sound and objective manner, automated
as much as possible.  We believe that many of the advanced tools
needed for this task already exist in the various fields of computer science
and statistics, including artificial intelligence (AI) and machine learning
(ML) techniques.  The specific challenge we have to address is the analysis of
data from a large digital sky survey, which is described below.  The 
techniques we are exploring are rather general, and should find many
applications well outside our immediate scientific target, including other
digital sky surveys, and indeed in virtually every data-intensive field. 

We have already applied some simple clustering analysis methods, both
supervised and partly-interactive, and we describe some of these scientific
applications below.  We have also started to explore unsupervised clustering
analysis on large data sets, including various Bayesian inference and cluster
analysis tools.  This goes beyond the mere visualisation of, and assistance
with, handling of huge data sets:  these may be software tools capable of 
{\it independent or cooperative discoveries}, and their application may greatly
enhance the productivity of practicing scientists. 

The need for the new data exploration tools for vast (Terabyte-sized) data sets
goes beyond the considerations of efficiency, important as that is.  In many
cases, including our own applications, the data sets are expected to change
and grow over a period of time, as more or better data, calibrations, etc.,
come in.  This is {\it an entirely new concept of an astronomical data
catalog:} a downloadable, growing data base with which one interacts using
semi-intelligent or semi-autonomous software agents.  The tools we are trying
to develop would be made generic to this concept of hypercatalogs.  There is a
fusion of the data and the information tools, and it is that new ground which
we plan to explore further.

\bigskip
\centerline{\bf 2.~ The Digital Palomar Observatory Sky Survey (DPOSS)}
\medskip

The specific data set which we are trying to explore is a digital sky survey,
DPOSS$^1$.  It is derived from a major new photographic sky atlas, the Second
Palomar Sky Survey (POSS-II), which is now nearing completion$^{2,3}$.  It will 
cover the entire northern sky with 894 fields ($6.5^\circ$ square) at $5^\circ$
spacings, and no gaps in the sky coverage.  Plates are taken in three
photometric bands: 
IIIa-J + GG395, $\lambda_{eff} \sim 480$ nm;
IIIa-F + RG610, $\lambda_{eff} \sim 650$ nm; and
IV-N   + RG9,   $\lambda_{eff} \sim 850$ nm.
Typical limiting magnitudes reached are $B_J \sim 22.5$, $R_F \sim 20.8$, and
$I_N \sim 19.5$, i.e., $\sim 1^m - 1.5^m$ deeper than the POSS-I.  The image
quality is improved relative to the POSS-I, and is comparable to the southern
photographic sky surveys.

These plates are being digitized at STScI, using modified PDS scanners, with
15-micron (1.0 arcsec) pixels, in rasters of 23,040 square, giving $\sim 1$
GB/plate, or $\sim 3$ TB of pixel data total for the entire survey$^{4}$.
Preliminary astrometric solutions are good to $\sim 0.5$ arcsec, and will get
better soon.  There is a major ongoing effort at Caltech to process and
calibrate the scans, and catalog and classify all objects detected down to the
survey limit.  We are using SKICAT, a novel software system developed for this
purpose$^{5,6,7,8,9,10,11}$, which incorporates some standard astronomical
image processing packages, commercial Sybase DBMS, as well as a number of AI
and ML based modules. 

A particular strength of SKICAT is the star-galaxy classification, which uses
artificial induction decision tree techniques$^{5,6,7,8,9,10,11,12,13}$.  By 
using these methods, and
using superior CCD data to train the AI object classifiers, we are able to
achieve classification accuracy of 90\% or better down to $\sim 1^m$ above the
plate detection limit; traditional techniques achieve comparable accuracy
typically only $\sim 2^m$ above the detection limit.  This effectively triples
the number of usable objects for most scientific applications of these data,
since in most cases one wants either stellar objects or galaxies. 
Future technical developments include an improved treatment of very bright
and/or extended objects, optimization of the object measurements for
crowded regions (e.g., low Galactic latitudes), better structuring of the
catalog database for efficient access and manipulation, etc.

An extensive CCD calibration effort is now underway at the Palomar 60-inch
telescope, and we expect it to expand to other sites soon.  The data are
calibrated in the Gunn $gri$ system.  We obtain at least 2 CCD fields per sky
survey field, and sometimes more.  These CCD images are used both for magnitude
zero-point calibrations, and for training of automated star-galaxy classifiers.
In addition to the CCD calibrations, we use heavily smoothed sky measurements
from the plate scans themselves (after the object removal) to ``flatfield''
away the telescope vignetting affects and the individual plate emulsion 
sensitivity variations.

As a result, we have demonstrated an unprecedented photometric stability and
accuracy for this type of photographic plate material$^{14}$.  We have
performed tests using both CCD sequences and plate overlaps, and find that our
magnitude zero-points are stable to within a few percent, both across the
plates, both between adjacent plates, and across the individual plates. 
Typical r.m.s. in the magnitude zero-points between different plates is in the
range $0.015^m - 0.045^m$ in the $r$ band, and slightly worse in the $g$ band,
perhaps due to the larger color terms in the J/$g$ calibration.  Keeping the
systematic magnitude zero-point errors below 10\% is essential for many
scientific applications of these data.  Median random magnitude errors for
stellar objects in all three bands start around $0.05^m$ at the bright end, and
increase to 
$\sim 0.25^m$ at $g_{lim} \approx 22^m$,
$\sim 0.20^m$ at $r_{lim} \approx 21.5^m$, and
$\sim 0.25^m$ at $i_{lim} \approx 20^m$.
For galaxies, these errors are typically higher by about 50\% at a given
magnitude.

The resulting data product, the Palomar-Norris Sky Catalog (PNSC) will contain
all objects down to an equivalent limiting magnitude of $B_J \sim 22^m$, with
star-galaxy classification accurate to 90\% or better down to $B_J \sim 21^m$. 
The PNSC is expected to contain $> 50$ million galaxies, and $> 2$ billion
stars (limited by the crowding at low Galactic latitudes), including $\sim
10^5$ quasars.  We note that the size of the DPOSS data set, in terms of the
bits, numbers of sources, and resolution elements, is $\approx 1,000 ~\times$
the entire IRAS data set, and is $\approx 0.1 ~\times$ the anticipated Sloan
Digital Sky Survey$^{15}$ data set.

\bigskip
\centerline{\bf 3.~ Examples of Scientific Applications}
\medskip

This large new database should be a fertile ground for numerous scientific
investigations, for years to come.  The nature of the data dictates its uses:
these images are not very deep by modern standards, but they do cover a very
large solid angle, and do so relatively uniformly.  In addition to the obvious
applications such as large-scale optical identifications of sources from other
wavelengths (e.g., radio, x-ray, IR), there are two general kinds of studies
which can be pursued very effectively with data sets of this size:
First, there are statistical astronomy studies, where the sheer large numbers
of detected sources tighten the statistical errors and allow for more model
parameters to be constrained meaningfully by the data. 
Second, and perhaps most interesting, there are searches for rare types of
objects.  For example, at intermediate Galactic latitudes, about one in a
million stellar objects down to $r \approx 19.5^m$ is a quasar at $z > 4$, yet
we can find such quasars very efficiently. 

We have already started a number of scientific projects using DPOSS, which also
serve as scientific verification tests of the data, and which have helped us
catch some errors and improve and control the data quality.  For example,
galaxy counts and colors in 3 bands from DPOSS can serve as a baseline for
deeper galaxy counts and a consistency check for galaxy evolution models$^{14}$.
Galaxy correlation functions and power spectra of galaxy clustering provide
useful constraints on scenarios of large scale structure formation$^{16}$.  
We are now also starting to explore the correlations of our galaxy counts with
IRAS and DIRBE infrared cirrus maps, in order to generate Galactic extinction
maps superior to those now commonly used. Other extragalactic projects now
planned include a catalog of $\sim 10^5$ brightest galaxies in the northern
sky, with a quantitative surface photometry and morphological information,
automated searches for low surface brightness galaxies, an archival search for
supernov\ae\ from plate overlaps, derivation of photometric redshift estimators
for galaxies, etc. 

Two of our ongoing projects illustrate well the need for advanced and automated
data exploration techniques, and clustering analysis in particular: an
automated search for clusters and groups of galaxies, and a search for
high-redshift quasars.  We describe these in turn. 

We are now starting a project to generate an objectively defined, statistically
well defined catalog of rich clusters of galaxies.  There are many cosmological
uses for rich clusters of galaxies$^{17}$.  They provide useful constraints for
theories of large-scale structure formation and evolution, and represent
valuable samples of galaxies to study their evolution in dense environments. 
Studies of the cluster two-point correlation function are a powerful probe of
large-scale structure, and the scenarios of its formation.  Correlations
between optically and x-ray selected clusters are also of considerable
scientific interest.  Most of the studies to date have been limited by the
statistical quality of the available cluster samples.  For instance, the
subjective nature of the most commonly used Abell catalog$^{18}$ has been
widely recognized as its major limitation.  Still, many far-reaching
cosmological conclusions have been drawn from it.  There is thus a real need to
generate well-defined, objective catalogs of galaxy clusters and groups, with
well understood selection criteria and completeness. 

Uniform DPOSS data and well-defined algorithms can produce a vastly superior,
quantifiable catalog of northern galaxy clusters and compact groups$^{19}$,
reaching considerably deeper than the Abell catalog.  The DPOSS has two main
advantages over previous surveys: it goes a magnitude fainter than POSS-I, and
allows us to use color information in selecting our candidates$^{33}$.  The
digitization and cataloging of objects allow us to search for clusters in a
consistent, objective way, producing statistically sound and well understood
samples of galaxy clusters. With our DPOSS data, we have already shown that we
can find all believable clusters which Abell found, and many more comparable
ones which he missed.  Furthermore, with the superior POSS-II data, we should
be able to find rich clusters at higher redshifts, perhaps up to $z \sim 0.5$,
with many at $z \sim 0.2 - 0.3$.  We estimate that eventually we will have a
catalog of as many as 20,000 rich clusters of galaxies at high Galactic
latitudes in the northern sky. 

The technical challenge here is to separate statistically signifficant clusters
of galaxies (which are defined in the 3-d space), from the overall $projected$
distribution of stars and galaxies on the sky.  The problem is greatly
complicated by two issues:  First, galaxies are not distributed in a random
Poissonian manner, but are highly correlated spatially; essentially, the
projected distribution of galaxies on the sky is $\sim 1/f$ noise.  The trick
is to isolate physical clusters from chance density peaks in this correlated
noise background and foreground.  The second problem is that clusters
themselves are inherently ill-defined, without obvious boundaries, and
frequently consisting of multiple density clumps (reflecting their dynamical
youth, still merging), and spanning a range of richness.  Thus, simple
density-peak schemes are rarely adequate for this task, and a more
sophisticated approach is called for. 

Our cluster selection algorithm works as follows.  We use the objects
classified as galaxies in DPOSS catalogs, and limit the samples at $r = 19.6^m$
(roughly equivalent to $B \approx 20.5^m - 21^m$) in order to maintain the
accuracy of object classifications at a $> 90$\% level.  This typically yields
about 50,000 galaxies per DPOSS field.  The next step is the color selection of
the candidate cluster galaxies.  We use cuts in the color space to select a
locus of probable early-type galaxies as they should better delineate
high-density regions.  We then use the adaptive kernel method$^{20}$ to create
the surface density maps.  A major advantage of this method is that it uses a
two-step process which significantly smooths the low density regions, and at
the same time keeps the high density peaks almost untouched.  This is superior
to the usual binning plus smoothing approach used by many previous
investigations. 

Next we evaluate the statistical significance of the density peaks using a
bootstrap technique to generate the a statistical significance map associated
with a given surface density map.  Density peaks are found using a version of
the FOCAS peak-finding algorithm.  We set our threshold at a 4-$\sigma$
level, where we successfully recover all of the known Abell clusters of
richness class 0 and higher, and also a large number of new cluster candidates
which were apparently missed by Abell.  We typically find of the order of 1 --
1.5 cluster candidates per square degree.  The important point is that we can
quantify our completeness and model our contamination. 

The same techniques we use to search for galaxy clusters can then be applied to
our star catalogs, in an objective and automated search for sparse globulars in
the Galactic halo, tidal disruption tails of former globular star clusters, and
possibly even new dwarf spheroidal galaxies in the Local Group. 

Another ongoing project is a survey for luminous quasars at $z > 4$.  Quasars
at $z > 4$ are valuable probes of the early universe, galaxy formation, and the
physics and evolution of the intergalactic medium at large redshifts.  They
probably mark sites of the earliest galaxy formation.

The technical challenge here is to isolate those objects which are
morphologically indistinguishable from stars (i.e., have the PSF shape) from a
vastly greater number of actual stars at comparable flux levels.  We expect
that about 400 or so of these quasars are detectable in the entire DPOSS data
base, which will also contain $\sim 2 \times 10^9$ stars.  Even at the high to
moderate Galactic latitudes where the contamination by foreground stars is
reduced, we find about one $z > 4$ quasar per million stars.  Thus, we need an
algorithm which can separate our ``signal'' from a foreground which is a
million times higher.  In practice, all one ever gets are quasar candidates,
which then must be checked spectroscopically.  An acceptable rate of false
positives may be about 10:1, but certainly not $10^6$:1. 

Given the lack of distinguishing morphological information for these objects,
we turn to a different portion of the parameter space, viz., the colors.  The
continuum drop across the Ly$\alpha$ line gives these objects a distinctive
color signature:  extremely red in $(g-r)$, yet blue in $(r-i)$, thus standing
away from the stellar sequence in the color space.  Traditionally, the major
contaminant in this type of work are red galaxies, which could mimic the
quasar colors.  This is where the morphological information comes back in.
Our superior star-galaxy classification leads to a manageable number of
color-selected candidates, and an efficient spectroscopic follow-up. 

As of this writing, about 30 new $z > 4$ quasars have been discovered.  Our
initial results$^{21,22,23,24}$ are the best estimates to date of the bright
end of the quasar luminosity function at $z > 4$.  We have thus verified the
decline in the comoving number density of bright quasars at $z > 4$.  There are
also some intriguing hints of possible primordial large-scale structure as
marked by these quasars.  However, much more data is needed to check this
potentially very exciting cosmological result. 

We can also search for stars with unusual colors or variability.  We have
started a search for stars at the bottom of the main sequence and field brown
dwarf candidates, using colors: anything with $(r-i) > 2.5$ should be
interesting.  At high Galactic latitudes, about one star in a few million is
that red, down to the conservative limit used so far ($r < 19.5^m$).  Such a
survey can be made much more powerful with the addition of IR data, such as
the Two-Micron All-Sky Survey (2MASS)$^{25}$.

\bigskip
\centerline{\bf 4.~ New Directions: Usupervised Classification and}
\centerline{\bf the Global Exploration of the DPOSS Data Space}
\medskip

A clustering algorithm, in general, is given input consisting of objects for
each of which there is a set of measurements$^{26}$.  Hence, objects are points
or vectors in some multidimensional data parameter space.  The algorithm is 
required to decide which groups of objects belong together in a cluster in this
parameter space.  If a distance measure is definable over the space, an
algorithm can use this measure as a basis of a set of clusters which minimize
intercluster distances but maximize intercluster distances.  In general, a
distance measure may not be defined and the problem is cast in a more general
form: Hypothesize a set of models governing the observation space, and find the
appropriate parameters of the models to best fit the data, or equivalently the
probability distribution(s) that underlay the observed data. 

The problem of fitting distributions to data is a difficult and computationally
expensive one.  Given $k$-dimensional data vectors (assuming that there are $k$
measurements for each object), the algorithm needs to decide the following: (1)
how many classes (clusters) are present in the data, and (2) how to describe
each of these classes. The first item involves searching over a large number of
possible partitions of the data.  Each partition specifies different regions
across which the data is quantitatively and qualitatively different.
Specifically, that means in each of the $J$ blocks of a chosen partition, the
data is governed by a given distribution (class model).  The second item
involves a difficult search over possible models and over the large space of
possible parameters (degrees of freedom) of a model.  For example, if one
hypothesizes that the data is normally distributed within each class, then the
parameters would be the mean vector and covariance matrix of the
multi-dimensional Gaussian distribution. 

Most clustering approaches leave it to the user to guess the number of classes
$J$; however, the most general ones have a method for estimating this number. 
In terms of models, most algorithms assume that observations within a cluster
(class) are normally distributed. This is motivated by theoretical results
stating that {\it any} distribution can be approximated fairly accurately with
a mixture of Gaussians (assuming one is allowed to use as many Gaussian bumps
as needed). 

AutoClass$^{27}$ is an unsupervised learning algorithm that fits user-specified
probability distribution models to a set of examples represented as feature
vectors.  Object classes are represented as particular parametrizations of the
models; typically, multi-dimensional Gaussian distributions are used. In order
to decide how many classes there are, and what parameters to set for each,
AutoClass uses a Bayesian strategy. This means that, for a data set $D$ deemed
to be in one class modeled by model $M_i$ with parameter vector $\theta$, the
program tries to find $\theta$ such that: 
$ \hbox{Prob}(\theta|D,M_i) $
is maximized. This quantity is actually computed by noting that, by Bayes rule, 
$$ 
\hbox{Prob}(\theta|D,M_i) =  
{{\hbox{Prob}(D|\theta,M_i) \hbox{Prob}(D) }
\over 
{\hbox{Prob}(D|M_i)\hbox{Prob}(M_i) }}
$$
The $\hbox{Prob}(D|\theta,M_i)$ is easily computed as a degree of fit, while
the other terms represent {\it prior probabilities} that may be estimated
or provided based on knowledge of the problem.  For example, the prior
expectation of a scientist might favour model $M_1$ (an exponential) over model
$M_2$ (say, a Gaussian). Hence, the prior on $\hbox{Prob}(M_i)$ can be set
accordingly.  In a similar fashion, AutoClass tries to find the most probable
number of classes $J$ by comparing the likelihoods of the fits for different
numbers of classes. That is, find the $J$ which maximizes 
$ \hbox{Prob}(J|M_i,\theta, D) $. 
An advantage to this scheme is that once models are found, one can obtain for
new objects (observations) membership probabilities in the formed classes. 

Thus we investigate the possibility of finding natural (data-based) partitions
of the attribute spaces which show high correlations between the plate-measured
attribute space, and the CCD-based attribute space, or a high degree of
separation between expected classes such as stars versus galaxies, spirals
versus ellipticals, or galaxies of different concentrations.  These partitions
of the data may be used for investigations of unusual regions of the attribute
space, and may even lead to the discovery of previously unknown classes of
objects.  We used a small subset of data from DPOSS, with only 7 attributes per
object.  We intentionally did not use several legitimate attributes like colors,
mean surface brightness and concentration index, which are available in our
catalogs, because at this point they can help us understand the association
between the classes which come out from the experiment and the large scale
distribution of galaxies.  Also, the classification is not given to the
algorithm but is only used to judge its performance. 

In our preliminary experiments using Bayesian clustering algorithms to classify
objects present in the DPOSS$^{28,29}$ we found that AutoClass was able to form
several sensible categories from a few simple attributes of the object images,
separating the data into four recognizable and astronomically meaningful
classes.  The results were robust and repeatable from field to field, and not a
function of magnitude.  The four classes found can be identified with stars
(s), galaxies with a bright core (g1), galaxies without a bright core (g2), and
stars with fuzz around them (sf). Thus, the object classes found by AutoClass 
are astronomically meaningful -- even though the program itself does not know
about stars, galaxies and such. Moreover, the two morphologically distinct
classes of galaxies populate different regions of the data space, and have
systematically different colors and concentration indices, even though
AutoClass was not given the color information!  Thus, {\it the program has
found astrophysically meaningful distinction between these classes of objects,
which is then confirmed by independent data.} 

One scientifically interesting lead is the physical nature of the objects
classified as ``fuzzy stars''.  Our preliminary follow-up indicates that a
large number of them may be previously unrecognized active galactic nuclei,
e.g., Seyfert galaxies.  This opens an exciting possibility of the automated
discovery of active galaxies by many thousands, which could lead to a range of
astrophysical applications: from an improved luminosity function of active
galactic nuclei, to their use as tracers of very large structure, to the
origins of the x-ray background.  It is very likely that other astrophysical
classes of objects can be separated once we use additional information (e.g.,
the colors).  One intriguing possibility would be extremely compact nucleated
dwarf galaxies, which may be the progeny of the mysterious starforming dwarf
galaxies at redshifts $z \sim 0.3 - 0.5$, the notorious faint blue galaxies
seen in deep galaxy counts.  Discovery of such a population in the DPOSS data
base would be of a great interest for cosmology. 

A critical point in constructing scientifically useful object catalogs is the
star/galaxy separation.  Various supervised classification schemes can be used
for this task$^{12,30,31}$.  However, a more difficult problem is to provide at
least rough morphological types for the galaxies detected, in a systematic and
objective way, without visual inspection of the images, which is obviously
impractical. We have thus started to explore new clustering analysis and
unsupervised classification techniques to try to separate astronomically
meaningful morphological types on the basis of the data themselves, rather than
some preconceived scheme. 

Similar approaches can be made to the automated pattern recognition and
classification of objects in the pixel domain.  One is then solving the
following inference problem: given the pixel data, what is the most likely
hypothesis regarding the physical features which yielded these particular data?
In particular, for a given new object, we are interested in composite
hypotheses concerning the presence or absence of a new class.  The problem is
ill-posed; the pixel data are inherently ambiguous due to the inevitable
resolution limitations and noise. A further complication arises because the
objects of interest and the surrounding sky background can vary substantially
in manners that are not easy to predict.  The natural framework for solving
problems of this nature is a probabilistic one: how to find the composite
hypotheses which accumulate the greatest probability mass conditioned on the
data?  The actual computation of posterior probabilities is relatively trivial
given a model. 

A more sophisticated approach is to model this in a Bayesian manner and
complement the widely-used squared-error goodness-of-fit metric with a penalty
term to reflect prior bias in terms of expected appearance of the object. For
example, one might try to adjust the boundaries of the segmented pixels to
minimize the following objective function: 
$$ E = \prod_{i=1}^N \sum_{k=1}^K p(x_i|\theta_k,s_k)p(\theta_k,s_k) $$
where $N$ is the total number of pixels, $i$ is an index over the pixels, $K$
is the total number of segments, $s_k$ is a particular segment and $\theta_k$
are the parameters for $s_k$.  The first term, $p(x_i|\theta_k,s_k)$, is the
likelihood of the pixel data given a particular set of segment parameters
$\theta_k$ and segment $s_k$. The segment parameters $\theta_k$ could include
the area, shape characteristics such as eccentricity, mean intensity, moments,
and so forth. The second term $p(\theta_k,s_k) = p(\theta_k|s_k)p(s_k)$
reflects our {\it a priori} expectations on the parameters themselves, e.g. a
particular combination of intensity and shape is highly unlikely to occur. More
complicated models would model the interaction between the segmented regions;
e.g. their relative positions.  The use of a prior knowledge to guide the
search (where the problem is modelled probabilistically using the Bayesian
formalism) is relatively novel in this type of application. 

Direct application of these techniques to the DPOSS database also represents a
novel and powerful form of quality control for our data products, as
multidimensional clustering can reveal subtle mismatch patterns between
individual sky survey fields, e.g., due to otherwise imperceptible calibration
variations.  This would apply to virtually any other digital sky survey or
other patchwise collated data sets. 

\bigskip
\centerline{\bf 5.~ Concluding Remarks}
\medskip

These algorithms may be also used for an objective discovery of clusters of
stars or galaxies in physical space, by utilizing the full information
available in the catalog.  For example, in addition to the clustering on the
sky itself, galaxies or stars belonging to a physical cluster have a well
defined apparent luminosity function, and correlations between colors,
magnitudes, surface brightness (for galaxies), etc., so that clusters in
physical space may be even more prominent in the suitably defined parameter
space.  This is indeed the concept behind the matched-filter approach$^{32}$,
but it is more general.

Likewise, these methods can be used in an automated search for high-redshift
quasars, using their distinct color signature in the color parameter space.  
We have already demonstrated this using simpler, semi-interactive techniques
described above.  However, a purely automated, objective selection has 
obvious statistical advantages, for both quasar and cluster searches.  

Perhaps the most exciting scientific prospects in this kind of studies involve
serendipity: with a data set as large as DPOSS, there is even a real
possibility of discovering some heretofore unknown types of objects or
phenomena, whose rarity would have made them escape astronomers' notice so far.

These, and other studies now started or planned, should produce many interesting
and useful new results in the years to come.  Availability of large data sets
such as DPOSS over the Net or through other suitable mechanisms would also
enable astronomers and their students anywhere, even if they are far from the
major research centers or without an access to large telescopes, to do some
first-rate observational science.  This new abundance of good data may 
profoundly change the sociology of astronomy, which is still dominated by a
few major research centers.

These techniques are clearly and directly applicable to a wide variety of
astronomical imaging applications, especially sky surveys of any sort, e.g.,
IRAS, Rosat, 2MASS, etc.  In addition to the efficient analysis of vast amounts
of new data, these techniques can also be used to explore existing data
archives, and have a potential of revolutionizing archival research (e.g., the
HST archive, reanalysis of IRAS or Rosat data, etc.).  This great universality
should attract a very broad constituency of science users, probably with a
multitude of applications which have never occurred to us.

\bigskip
\centerline{\bf Acknowledgements}
\medskip

This work was supported in part by the funds from NASA, the Norris Foundation,
and the NSF PYI award AST-9157412.  We acknowledge the efforts of the POSS-II
team at Palomar, the digitization team at STScI.  N.~Weir and U.~Fayyad made
important initial contributions to this project.  We also thank J.~Kennefick,
J.~Darling, and V.~Desai for their contributions to the quasar search project.
The DPOSS work at Caltech is a part of the CRONA international collaboration.

\bigskip
\bigskip
\centerline{\bf References}
\medskip
\parskip=0pt

\pp  1.  
Djorgovski, S.G., de Carvalho, R.R., Gal, R., Pahre, M.A., Scaramella, R., and
Longo, G. 
``Cataloging of the Digitized POSS-II, and Some Initial Scientific Results From
It'', 
in {\sl New Horizons From Multi-Wavelength Sky Surveys}, Proc. IAU Symp. \#179,
B.~McLean {\it et al.} (eds.), in press, Dordrecht: Kluwer, 1997. 

\pp  2. 
Reid, I. N., \etal,
``The Second Palomar Sky Survey'',
{\it Publ. Astron. Soc. Pacific} {\bf 103}, 661, 1991.

\pp  3. Reid, I.N., and Djorgovski, S., 
``The Second Palomar Sky Survey'',
in {\sl Sky Surveys: Protostars to Protogalaxies}, B.T. Soifer (ed.), 
{\it A.S.P.~Conf.~Ser.} {\bf 43}, 125, 1993.

\pp  4. 
Lasker, B., Doggett, J., McLean, B., Postman, M., Sturch, C.,
Djorgovski, S., de Carvalho, R., and Reid, I.N. 
``The Palomar -- STScI Digitized Sky Survey:  Preliminary Data Availability'',
in {\sl Astronomical Data Analysis Software and Systems V}, eds. G. Jacoby and
J. Barnes, {\it A.S.P.~Conf.~Ser.} {\bf 101}, 88, 1996. 

\pp  5.
Weir, N., Fayyad, U., Djorgovski, S.G., and Roden, J.
``The SKICAT System for Processing and Analysing Digital Imaging Sky Surveys'',
{\it Publ. Astron. Soc. Pacific} {\bf 107}, 1243, 1995.

\pp  6.  
Fayyad, U.M., Weir, N., and Djorgovski, S., 
``SKICAT:  A Machine Learning System for Automated Cataloging of Large 
Scale Sky Surveys'',
in {\sl Proc. Tenth Intl. Conf. on Machine Learning},
p.~112.  San Mateo: Morgan Kaufmann Publ., 1993.

\pp  7. 
Djorgovski, S., Weir, N., and Fayyad, U., 
``Processing and Analysis of the Palomar -- STScI Digital Sky Survey Using a
Novel Software Technology'', 
in {\sl Astronomical Data Analysis Software and Systems III},
D. Crabtree, R. Hanisch, and J. Barnes (eds.),
{\it A.S.P.~Conf.~Ser.} {\bf 61}, 195, 1994.

\pp  8.  
Fayyad, U.M., Djorgovski, S.G., and Weir, N.,
``Cataloging Sky Objects using SKICAT'',
in {\sl Advances in Knowledge Discovery and Data Mining}, 
U. Fayyad, G. Piatetsky-Shapiro, P. Smyth, and R. Uthurusamy (eds.), 
Boston: AAAI/MIT Press, p.~471, 1996.

\pp  9.  
Weir, N., Fayyad, U., Djorgovski, S., Roden, J., and Rouquette, N., 
``SKICAT:  A Cataloging and Analysis Tool for Wide Field Imaging Surveys'' 
in {\sl Astronomical Data Analysis Software and Systems III},
R. Hanisch, R. Brissenden, and J. Barnes (eds.),
{\it A.S.P.~Conf.~Ser.} {\bf 52}, 39, 1996.

\pp 10.  
Weir, N., Djorgovski, S., Fayyad, U., Smith, J.D., and Roden, J.,
``Cataloging the Northern Sky Using a New Generation of Software Technology'', 
in {\sl Astronomy From Wide-Field Imaging}, Proc. IAU Symp. \#161, 
H. MacGillivray {\it et al.} (eds.), 
p.~205, Dordrecht: Kluwer, 1994. 

\pp 11.  
Fayyad, U., Smyth, P., Weir, N., and Djorgovski, S.,
``Automated Analysis and Exploration of Image Databases: Results, Progress, and
Challenges'', 
{\it J.~Intel.~Inf.~Sys.} {\bf 4}, 7, 1995.

\pp 12.  
Weir, N., Fayyad, U., and Djorgovski, S.,
``Automated Star/Galaxy Classification for Digitized POSS-II'',
{\it Astron.~J.} {\bf 109}, 2401, 1995.

\pp 13.  
Fayyad, U., Doyle, R., Weir, N., and Djorgovski, S.,
``Applying Machine Learning Classification Techniques to Automate Sky
Object Cataloguing'' 
in {\sl Proc. Intl. Space Year Conf. on Earth and Space Science Information
Systems}, AIP Conf. Proc. {\bf 283}, A. Zygielbaum (ed.), p.~405, New York:
AIP, 1993. 

\pp 14.  
Weir, N., Djorgovski, S., and Fayyad, U.,
``Initial Galaxy Counts From Digitized POSS-II'',
{\it Astron.~J.} {\bf 110}, 1, 1995.

\pp 15. 
Gunn, J. E., and Knapp, G.R.
``The Sloan Digital Sky Survey'',
in {\sl Sky Surveys: Protostars to Protogalaxies}, B.T. Soifer (ed.), 
{\it A.S.P.~Conf.~Ser.} {\bf 43}, 267, 1993.

\pp 16.  Brainerd, T., de Carvalho, R., and Djorgovski, S.,
``Clustering of Galaxies in the Digitized Palomar Observatory Sky Survey:
Preliminary Results'', 
{\it Bull. Am. Ast. Soc.} {\bf 27}, 1364, 1995.

\pp 17.  Bahcall, N.~A., 
``Large-Scale Structure in the Universe Indicated by Galaxy Clusters'',
{\it Ann. Rev. Astron. Astrophys.} {\bf 26}, 631, 1995.

\pp 18.  Abell, G. O., 
``The Distribution of Rich Clusters of Galaxies'',
{\it Astrophys.~J.~Suppl.} {\bf 3}, 211, 1958.

\pp 19.  
de Carvalho, R.R., Djorgovski, S.G., Pahre, M.A., Gal, R.R., Gray, A., and 
Roden, J.,
``Towards an Objectively Defined Catalog of Galaxy Clusters from the
Digitized POSS-II'',
in {\sl Wide Field Spectroscopy}, eds. E.~Kontizas {\it et al.}, p.~285. 
Dordrecht: Kluwer, 1997.

\pp 20. 
Silverman, B.W.,
{\sl Density Estimation for Statistics and Data Analysis},
London: Chapman \& Hall, 1986.

\pp 21.  
Kennefick, J.D., de Carvalho, R.R., Djorgovski, S.G., Wilber, M.M., Dickson,
E.S., Weir, N., Fayyad, U., and Roden, J.,
``The Discovery of Five Quasars at $z > 4$ Using the Second Palomar Sky
Survey'',
{\it Astron.~J.} {\bf 110}, 78, 1995.

\pp 22.  
Kennefick, J.D., Djorgovski, S.G., and de Carvalho, R.R.,
``The Luminosity Function of $z > 4$ Quasars from the Second Palomar Sky
Survey'',
{\it Astron.~J.} {\bf 110}, 2553, 1995.

\pp 23.  
Kennefick, J.D., Djorgovski, S.G., and de Carvalho, R.R.,
``The Space Density of $z > 4$ Quasars from the Second Palomar Sky Survey'', 
in {\sl Wide Field Spectroscopy}, eds. E.~Kontizas {\it et al.}, p.~381,
Dordrecht: Kluwer, 1997.

\pp 24.  
Kennefick, J.D., Darling, J.K., Djorgovski, S.G., and de Carvalho R.R.,
``The Luminosity Function of $z > 4$ Quasars From the Second Palomar Sky
Survey'',
in {\sl Young Galaxies and QSO Absorption-Line Systems}, eds.
S. Viegas, R. Gruenwald, and R. de Carvalho, 
{\it A.S.P.~Conf.~Ser.} {\bf 114}, 95, 1997.

\pp 25. 
Skrutskie, M., \etal,
``The Two-Micron All-Sky Survey (2MASS): Overview and Status'',
in {\sl The Impact of Large Scale Near-IR Sky Surveys}, eds. F. Garzon \etal,
p.~25, Dordrecht: Kluwer, 1997.

\pp 26.  Duda, R., and Hart, P., 
{\sl Pattern Classification and Scene Analysis},
New York: John Wiley and Sons, 1973.

\pp 27.  Cheeseman, P., \etal, ``AutoClass'',
in {\sl Proc. Fifth Machine Learning Workshop}, p.54, 
San Mateo: Morgan Kaufmann Publ., 1988. 

\pp 28.  
de Carvalho, R., Djorgovski, S., Weir, N., Fayyad, U., Cherkauer, K., 
Roden, J., and Gray, A., 
``Clustering Analysis Algorithms and Their Applications to Digital POSS-II
Catalogs'' 
in {\sl Astronomical Data Analysis Software and Systems IV},
R. Shaw, H. Payne, and J. Hayes (eds.),
{\it A.S.P.~Conf.~Ser.} {\bf 77}, 272, 1995.

\pp 29.  
Yoo, J., Gray, A., Roden, J., Fayyad, U., de Carvalho, R., and Djorgovski, S.,
``Analysis of Digital POSS-II Catalogs Using Hierarchical Unsupervised
Learning Algorithms'' 
in {\sl Astronomical Data Analysis Software and Systems V},
G. Jacoby and J. Barnes (eds.),
{\it A.S.P.~Conf.~Ser.} {\bf 101}, 41, 1995.

\pp 30. 
Odewahn, S.C., Stockwell, E., Pennington, R., Humphreys, R., and Zumach, W.,
``Automated Star-Galaxy Discrimination With Neural Networks'',
{\it Astron.~J.} {\bf 103}, 318, 1992.

\pp 31. 
Odewahn, S.C., 
``Automated Classification of Astronomical Images'',
{\it Publ. Astron. Soc. Pacific} {\bf 107}, 770, 1995.

\pp 32. 
Postman, M., Lubin, L., Gunn, J.E., Oke, J.B., Hoessel, J., Schneider, D.,
and Christensen, J.,
``The Palomar Distant Cluster Survey. I. The Cluster Catalog'',
{\it Astron.~J.} {\bf 111}, 615, 1996.

\pp 33. 
Odewahn, S.C., and Aldering, G.,
``Galaxy Properties at the North Galactic Pole. I. Photometric Properties on
Large Spatial Scales'',
{\it Astron.~J.} {\bf 110}, 2009, 1995.

\bigskip
\bigskip
\bigskip

\centerline{\bf Figure Captions}
\bigskip
\bigskip

\noindent {\bf Figure 1.}~~
Projected surface density of color-selected galaxies detected in the DPOSS
field 475, created with the adaptive kernel algorithm.  Presence of large-scale
structure, including possible clusters, is evident.  Statistically significant
clusters are circled.  While the known Abell clusters are among the more
significant peaks, there are some as statistically significant clusters which
he missed.  The majority of the new candidates are less rich and/or fainter,
presumably extending to higher redshifts. 

\bigskip
\bigskip

\noindent {\bf Figure 2.}~~
A typical color-color diagram of objects morphologically classified as stellar
in a POSS-II field (dots), with some of our newly discovered (spectroscopically
confirmed) high-redshift quasars plotted as solid circles.  Only the stars in a
narrow magnitude interval $19^m < r < 19.5^m$ are plotted here, for clarity. 
Galactic stars occupy a well-defined, banana-shaped locus in this parameter
space, while the high-redshift quasars cluster to the lower right in this
diagram.  The principal contaminant in the quasar candidate selection are
misclassified galaxies, which sometimes have colors like those of quasars.

\bigskip
\bigskip

\noindent {\bf Figure 3.}~~
An example of unsupervised object classification applied to DPOSS data.  This
shows the four statistically distinct classes of objects found by AutoClass in
a very small subset of data from DPOSS.  A set of 9 morphological image
parameters was used.  Four examples are shown for each of the four classes
found.  With an astronomical hindsight, they are labeled stars (s; first row),
stars with a fuzz (sf; second row), early-type galaxies (g1; third row) and
late-type galaxies (g2; fourth row).  As an independent check, the two types of
galaxies also separate cleanly when color information is used -- even though
the program was not given this information!  This demonstrates the power of an
unsupervised classification algorithm in finding astrophysically meaningful
classes of objects in a large data set.

\vfill
\eject
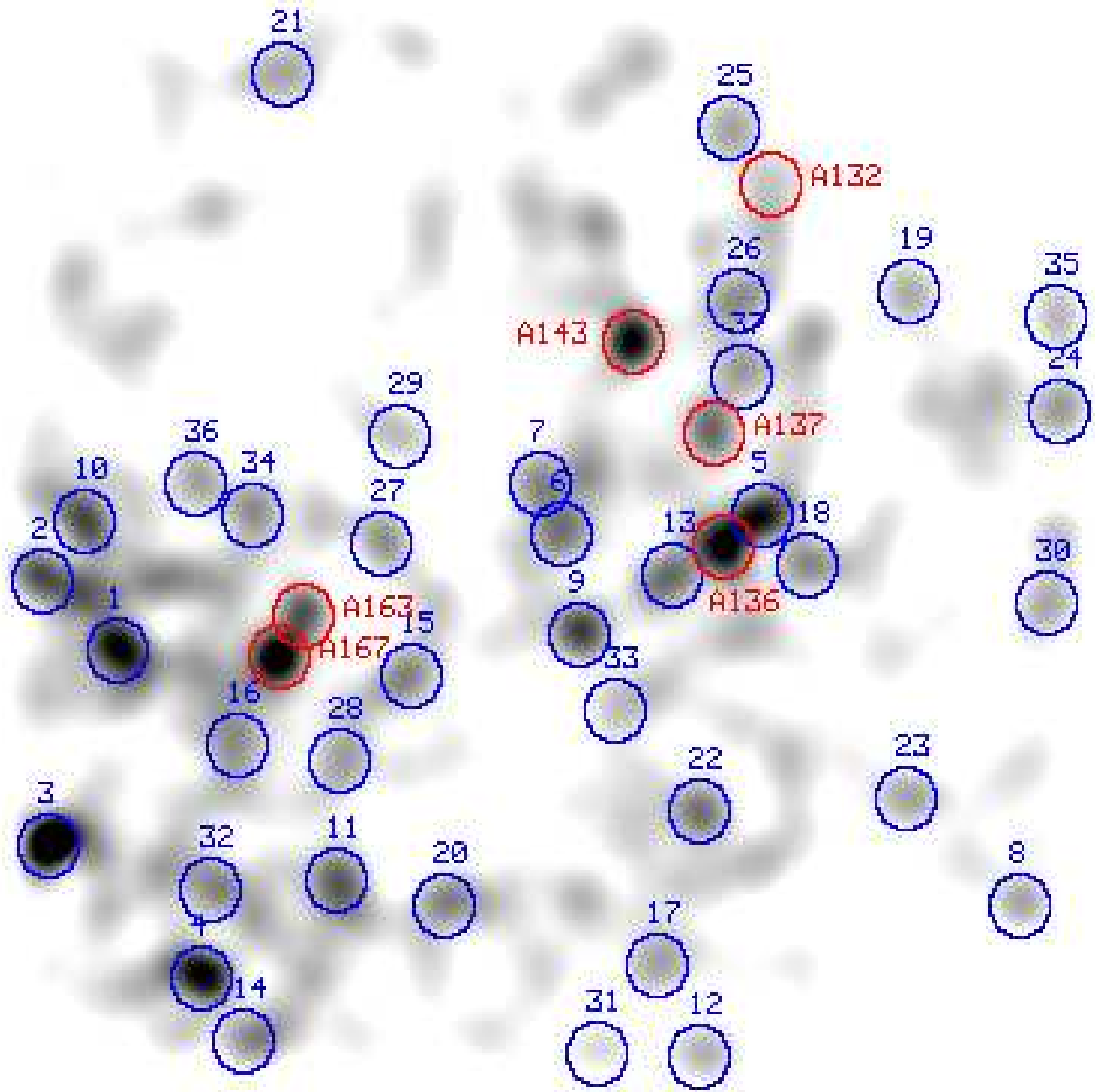
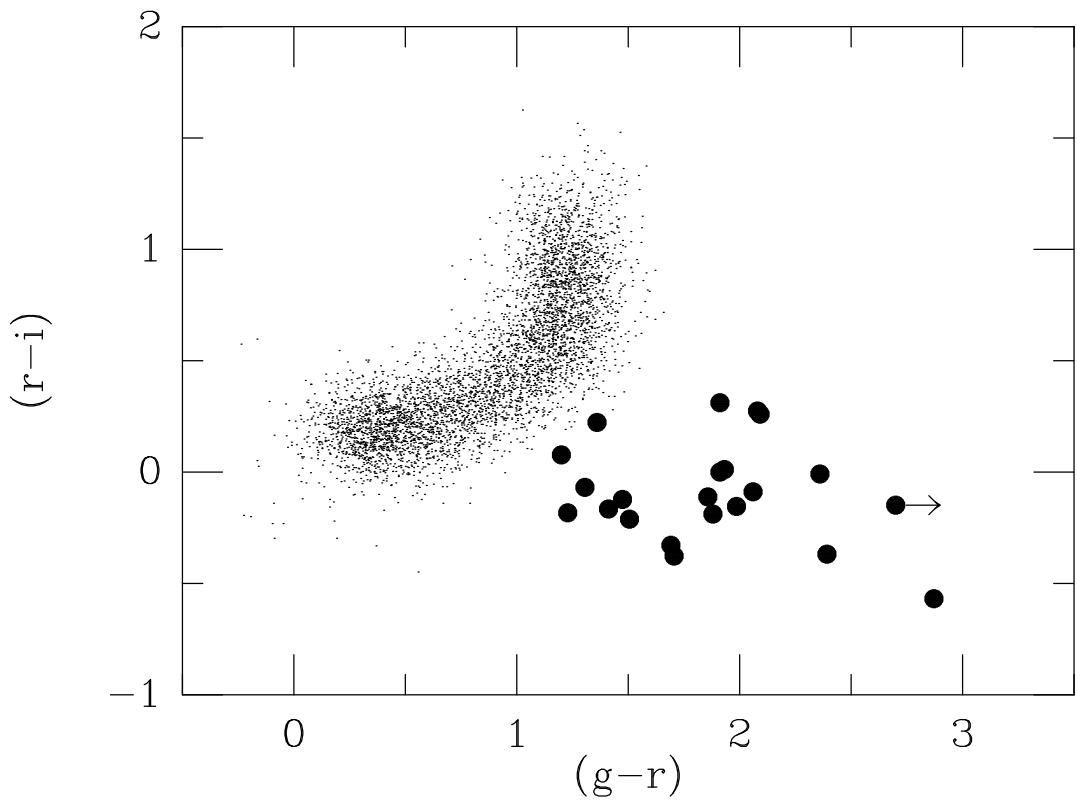
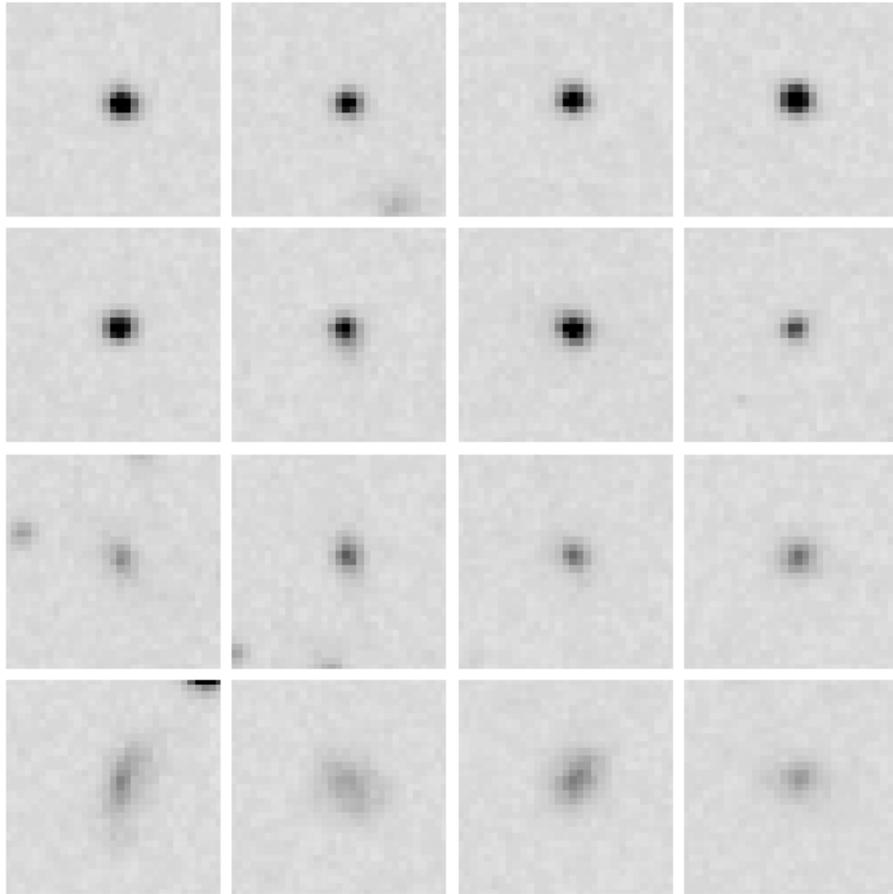
\end